\documentclass[sigconf]{acmart}

\usepackage{booktabs} 
\usepackage{todonotes}

\usepackage{tablefootnote}
\usepackage{threeparttable}
\usepackage{balance}

\copyrightyear{}
\acmYear{}
\setcopyright{none}

\acmConference[Copyright \textcopyright 2019 for this paper by its authors. Use permitted under Creative Commons License Attribution 4.0 International (CC BY 4.0). \\
In 7th International Workshop on News Recommendation and Analytics (INRA 2019), in conjunction with RecSys 2019, September 19, 2019, Copenhagen, Denmark.]{.}

\acmBooktitle{Proceedings of 7th International Workshop on News Recommendation and Analytics (INRA 2019), September 19, 2019, Copenhagen, Denmark}
\acmPrice{}
\acmDOI{}
\acmISBN{}

\begin{document}

\title[On the Importance of News Content Representation in Hybrid Neural...]{On the Importance of News Content Representation in Hybrid Neural Session-based Recommender Systems}

\author{Gabriel de Souza P. Moreira}
\affiliation{%
  \institution{CI\&T}
  \city{Campinas}
  \state{SP}
  \country{Brazil}
}
\additionalaffiliation{%
	\institution{Brazilian Aeronautics Institute of Technology}
	\city{S\~ao Jos\'e dos Campos}
	\state{SP}
	\country{Brazil}
}
\email{gspmoreira@gmail.com}

\author{Dietmar Jannach}
\affiliation{%
  \institution{University of Klagenfurt}
  \country{Klagenfurt, Austria}
}
\email{dietmar.jannach@aau.at}

\author{Adilson Marques da Cunha}
\affiliation{%
  \institution{Instituto Tecnol\'ogico de Aeron\'autica}
  \city{S\~ao Jos\'e dos Campos}
  \state{SP}
  \country{Brazil}
}
\email{cunha@ita.br}

\renewcommand{\shortauthors}{de Souza P. Moreira et al.}

\begin{abstract}
News recommender systems are designed to surface relevant information for online readers by personalizing their user experiences. A particular problem in that context is that online readers are often anonymous, which means that this personalization can only be based on the last few recorded interactions with the user, a setting named session-based recommendation. Another particularity of the news domain is that constantly fresh articles are published, which should be immediately considered for recommendation. To deal with this item cold-start problem, it is important to consider the actual content of items when recommending. Hybrid approaches are therefore often considered as the method of choice in such settings. In this work, we analyze the importance of considering content information in a hybrid neural news recommender system. We contrast content-aware and content-agnostic techniques and also explore the effects of using different content encodings. Experiments on two public datasets confirm the importance of adopting a hybrid approach. Furthermore, we show that the choice of the content encoding can have an impact on the resulting performance.

\end{abstract}
%


\maketitle



\section{Introduction \& Background}
Many of today's major media and news aggregator websites, including The New York Times \cite{spangher2015}, The Washington Post \cite{graff2015}, Google News \cite{das2007}, and Yahoo! News \cite{trevisiol2014cold}, provide automated reading recommendations for their users. News recommendation, while being one of the earliest application fields of recommenders, is often still considered a challenging problem for a many reasons \cite{karimi2018news}.

Among them, there are two types of cold-start problems.
First, there is the permanent \emph{item cold-start problem}. In the news domain, we have to deal with a constant stream of possibly thousands of new articles published each day \cite{spangher2015}. At the same time, these articles become outdated very quickly  \cite{das2007}.
Second, on many news sites, we have to deal with \emph{user cold-start}, when users are anonymous or not logged-in \cite{lin2014personalized,diez2016,li2011scene}, which means that personalization has to be based on a few observed interactions (e.g., clicks) of the user.

In many application domains of recommenders, collaborative filtering techniques, which only rely on observed preference patterns in a user community, have proven to be highly effective in the past. However, in the particular domain of news recommendation, the use of hybrid techniques, which also consider the actual content of a news item, have often shown to be preferable to deal with item cold-start, see e.g., \cite{chu2009personalized,liu2010personalized,li2011scene,rao2013personalized,lin2014personalized,li2014modeling,trevisiol2014cold,epure2017recommending}.

Likewise, to deal with user cold-start issues, \emph{session-based recommendation} techniques received more research interest in recent years. In these approaches, the provided recommendations are not based on long-term preference profiles, but solely on adapting recommendations according to the most recent observed interactions of the current user.

Technically, a number of algorithmic approaches can be applied for this problem, from rule-learning techniques, over nearest-neighbor schemes, to more complex sequence learning methods and deep learning approaches. For an overview see \cite{QuadranaetalCSUR2018}. Among the neural methods, Recurrent Neural Networks (RNN) are a natural choice for learning sequential models \cite{hidasi2016,Li2017narm}. Attention mechanisms have also been used for session-based recommendation \cite{Liu2018stamp}.

The goal of this work is to investigate two aspects of hybrid session-based news recommendation using neural networks. Our first goal is to understand the value of considering content information in a hybrid system. Second, we aim to investigate to what extent the choice of the mechanism for encoding the articles' textual content matters. To that purpose, we have made experiments with various encoding mechanisms, including unsupervised (like \emph{Latent Semantic Analysis} and \emph{doc2vec}) and supervised ones.
Our experiments were made using a realistic streaming-based evaluation protocol. The outcomes of our studies, which were based on two public datasets, confirm the usefulness of considering content information. However, the quality and detail of the content representation matters, which means that care of these aspects should be taken in practical settings. Second, we found that the specific document encoding \emph{can} makes a difference in recommendations quality, but sometimes those differences are small. Finally, we found that content-agnostic nearest-neighbor methods, which are considered highly competitive with RNN-based techniques in other scenarios \cite{ludewig2018evaluation,jannach2017recurrent}, were falling behind on different performance measures compared to the used neural approach.

\section{Methodology}
To conduct our experiments, we have implemented different instantiations of our deep learning \emph{meta-architecture} for news recommendation called \emph{CHAMELEON} \cite{moreira2018news,moreira2019contextual}.
The main component of the architecture is the  \emph{Next-Article Recommendation (NAR)} module, which processes various types of input features, including pre-trained \emph{Article Content Embeddings (ACE)} and contextual information about users (e.g., time, location, device) and items (e.g., recent popularity, recency). These inputs are provided for all clicks of a user observed in the current session to generate next-item recommendations based on an RNN (e.g., GRU, LSTM).

\begin{figure}[h]
	\includegraphics[width=0.50\columnwidth]{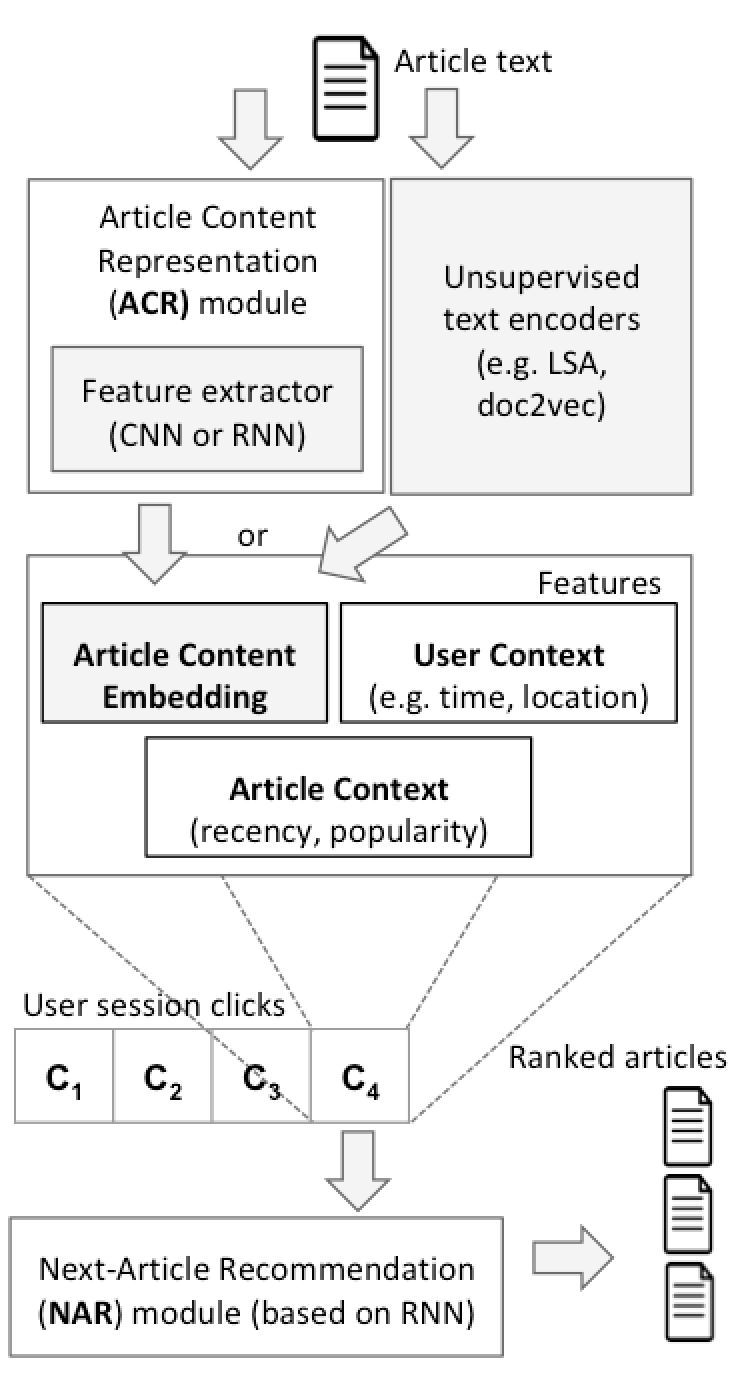}	\caption{A simplified overview of \emph{CHAMELEON}. The components for which we tested different variants are shaded.}
	\label{figure:chameleon}
\end{figure}

The \emph{ACEs} are produced by the \emph{Article Content Representation (ACR)} module. The input to the module is the article's text, represented as a sequence of word embeddings (e.g. using Word2Vec \cite{mikolov2013}), pre-trained on a large corpus. These embeddings are further processed by \emph{feature extractors}, which can be instantiated as Convolutional Neural Networks (CNN) or RNNs. The \emph{ACR module}'s neural network is trained in a supervised manner for a side task: to predict metadata attributes of an article, such as categories or tags. Figure~\ref{figure:chameleon} illustrates how the \emph{Article Content Embeddings} are used within \emph{CHAMELEON}'s processing chain to provide next-article recommendations.

In this work, we first analyzed the importance of considering article content information for recommendations. Second, we experimented with different techniques for textual content representation\footnote{As there were some very long articles, the text was truncated after the first 12 sentences, and concatenated with the title. \emph{Article Content Embeddings (ACE)} produced by the selected techniques were \emph{L2}-normalized to make the feature scale similar, but also to preserve high similarity scores for embeddings from similar articles.}, and investigated how they might affect recommendation quality. The different variants that were tested \footnote{We also experimented with Sequence \emph{Autoencoders GRU} (adapted from \emph{SA-LSTM} \cite{dai2015semi}) to extract textual features by reconstructing the sequence of input word embeddings, but this technique did not lead to better results than the other unsupervised methods.} are listed in Table~\ref{tab:feature-extraction}.

For the experiments, \emph{CHAMELEON}'s \emph{NAR} module took the following features as input, described in more detail in \cite{moreira2019contextual} \footnote{Note that the experiments reported here did not include the \emph{trainable Article ID} feature used in the experiments from \cite{moreira2019contextual}, which can lead to a slightly improved accuracy, but possibly reduces the differences observed between the content representations.}: (1) \emph{Article Content Embeddings} (generated by the different techniques presented in Table~\ref{tab:feature-extraction}), (2) article metadata (category and author\footnote{Article author and user city are available only for the \emph{Adressa} dataset.}), (3) article context (novelty and recency), (4) user context (city, region, country, device type, operational system, hour of the day, day of the week, referrer).

\begin{table}[h]
\begin{threeparttable}
\centering
\caption{Alternative content processing techniques.}
\vspace{-5pt}

\label{tab:feature-extraction}
\footnotesize
\begin{tabular}{p{1.2cm}lp{5.1cm}}
\hline
 \emph{Technique} &
 \emph{Input} &
 \emph{Description} \\
\hline
\emph{No-ACE} & \emph{None} & In this setting, no content representation is used as input. \\
\hline
\multicolumn{3}{p{6cm}}{\textbf{Supervised}}\\ \hline
\emph{CNN} & \emph{word2vec}\footnotemark & A 1D-CNN-based model trained to classify the articles' metadata (e.g., category). The architecture combines three CNNs, with window sizes of 3, 4, and 5 to model n-grams. The output of an intermediate layer is used as textual representation. For more details see \cite{moreira2018news,moreira2019contextual} \\
\emph{GRU} & \emph{word2vec} & Similar to the CNN-based version, a GRU layer is trained to classify metadata. The outputs of the GRU layer are max-pooled to generate representations.\\

\hline
\multicolumn{3}{p{6cm}}{\textbf{Unsupervised}}\\ \hline
\emph{LSA} & Raw text & Traditional \emph{Latent Semantic Analysis (LSA)} \cite{deerwester1990lsa}. We used a variation based on \emph{TF-IDF} vectors \cite{ramos2003using} and \emph{Truncated SVD} \cite{halko2011finding}. \\
\emph{W2V*TF-IDF} & \emph{word2vec} & \emph{TF-IDF} weighted word embeddings \cite{lilleberg2015support}, a technique to represent a piece of text as the average of its word embeddings weighted by \emph{TF-IDF} \cite{ramos2003using}. \\ 
\emph{doc2vec} & Raw text & Paragraph Vector (a.k.a \emph{doc2vec}) \cite{le2014distributed} learns fixed-length feature representations from variable-length pieces of texts, which are trained via the distributed memory and distributed bag of words models. \\
\hline
\end{tabular}
\end{threeparttable}
\vspace{-10pt}
\end{table}
\footnotetext{Portuguese: A pre-trained Word2Vec \cite{mikolov2013} \emph{skip-gram} model (300 dimensions) is available at \url{http://nilc.icmc.usp.br/embeddings}; and Norwegian: a \emph{skip-gram} model (100 dimensions) is available at \url{http://vectors.nlpl.eu/repository} (model \#100).}

\section{Experimental Setup}
We adopt a temporal offline evaluation method as proposed in \cite{moreira2018news,moreira2019contextual}, which simulates a streaming flow of new user interactions (clicks) and articles being published. Since in practical environments it is highly important to quickly react to incoming events \citep{ludmann2017recommending, kille2017clef,jugovac2018streamingrec}, the baseline recommender methods are constantly updated over time. \emph{CHAMELEON}'s \emph{NAR} module also supports online learning. 
The training process of \emph{CHAMELEON} emulates a streaming scenario with mini-batches, in which each user session is used for training only once. Such a scalable approach is different from other techniques, like \emph{GRU4Rec} \citep{hidasi2016}, which require training for some epochs on a larger set of past interactions to reach high accuracy.

\subsection{Evaluation Protocol} The evaluation process works as follows:  \vspace{2pt}

(1) The recommenders are continuously trained on user sessions ordered by time and grouped by hours. Every five hours, the recommenders are evaluated on sessions from the next hour. With this interval of five hours (not a divisor of 24 hours), we cover different hours of the day for evaluation. After the evaluation of the next hour was done, this hour is also considered for training, until the entire dataset is covered.\footnote{Our dataset consists of 16 days. We used the first 2 days to learn an initial model for the session-based algorithms and report the averaged measures after this warm-up.} 
Note that \emph{CHAMELEON}'s model is only updated after all events of the test hour are processed. This allows us to emulate a realistic production scenario  where the model is trained and deployed once an hour to serve recommendations for the next hour;

\vspace{2pt}
\noindent (2) For each session in the test set, we incrementally reveal one click after the other to the recommender, as done, e.g., in \cite{hidasi2016,quadrana2017personalizing};

\vspace{2pt}
\noindent (3) For each click to be predicted, we sample a random set containing 50 recommendable articles (the ones that received at least one click by any user in the preceding hour) that were \emph{not} viewed by the user in their session (negative samples) plus the true next article (positive sample), as done in \cite{Cremonesi2010} and \cite{koren2009}. We then evaluate the algorithms for the task of ranking those 51 items; and

\vspace{2pt}
\noindent (4) Given these rankings, standard information retrieval (top-n) metrics can be computed.

\subsection{Metrics}
As relevant quality factors from the news domain \cite{karimi2018news}, we considered accuracy, item coverage, and novelty. To determine the metrics, we took measurements at list length 10. As accuracy metrics, we used the \emph{Hit Rate (HR@n)}, which checks whether or not the true next item appears in the top-n ranked items, and the \emph{Mean Reciprocal Rank (MRR@n)}, a ranking metric that is sensitive to the position of the true next item. Both metrics are common when evaluating session-based recommendation algorithms \citep{hidasi2016,ludewig2018evaluation,jugovac2018streamingrec}.

Since it is sometimes important that a news recommender not only focuses on a small set of items, we also considered \emph{Item Coverage (COV@n)} as a quality criterion. We computed item coverage as the number of distinct articles that appeared in any top-n list divided by the number of recommendable articles \cite{jannach2015recommenders}. In our case, the recommendable articles are the ones viewed at least once in the last hour by any user. To measure novelty, we used the \emph{ESI-R@n} metric \cite{moreira2019contextual}, adapted from \cite{vargas2011rank, castells2015novelty, vargas2015thesis}. The metric is based on item popularity and returns higher values when long-tail items are among the top-n recommendations.

\subsection{Datasets}

We use two public datasets from news portals: 

\vspace{2pt}
\noindent (1) \emph{Globo.com} (\emph{G1}) dataset -  Globo.com is the most popular media company in Brazil. The dataset\footnote{\url{https://www.kaggle.com/gspmoreira/news-portal-user-interactions-by-globocom}} was collected at the G1 news portal, which has more than 80 million unique users and publishes over 100,000 new articles per month; and

\vspace{2pt}
\noindent (2) \emph{SmartMedia Adressa} - This dataset contains approximately 20 million page visits from a Norwegian news portal \cite{gulla2017adressa}. In our experiments we used its complete version\footnote{\url{http://reclab.idi.ntnu.no/dataset}}, which includes article text and click events of about 2 million users and 13,000 articles.

\vspace{2pt} Both datasets include the textual content of the news articles, article metadata (such as publishing date, category, and author), and logged user interactions (page views) with contextual information. Since we are focusing on session-based news recommendations and short-term users preferences, it is not necessary to train algorithms for long periods. Therefore, and because articles become outdated very quickly, we have selected all available user sessions from the first 16 days for both datasets for our experiments.

In a pre-processing step, like in \cite{ludewig2018evaluation, epure2017recommending, twardowski2016modelling}, we organized the data into sessions using a 30 minute threshold of inactivity as an indicator of a new session.
Sessions were then sorted by timestamp of their first click. From each session, we removed repeated clicks on the same article, as we are not focusing on the capability of algorithms to act as reminders as in \cite{LercheJannachEtAl2016}. Sessions with only one interaction are not suitable for next-click prediction and were discarded. Sessions with more than 20 interactions (stemming from \emph{outlier} users with an unusual behavior or from bots) were truncated.

The characteristics of the resulting pre-processed datasets are shown in Table~\ref{tab:datasets}. Coincidentally, the datasets are similar in many statistics, except for the total number of published articles, which is much higher for \emph{G1} than for the \textit{Adressa} dataset.

\begin{table}[h!t]
\centering
\caption{Statistics of the datasets used for the experiments.}
\label{tab:datasets}
\vspace{-5pt}
\footnotesize
\begin{tabular}{p{3cm}rr}
\hline
 & \emph{Globo.com (G1)}
 & \emph{Adressa} \\ \hline
Language  & Portuguese & Norwegian  \\ 
Period (days)  & 16 & 16 \\  
\# users   & 322,897 & 314,661  \\ 
\# sessions & 1,048,594 & 982,210 \\  
\# clicks  & 2,988,181 & 2,648,999 \\  
\# articles   & 46,033 & 13,820  \\ 
Avg. session length & 2.84  & 2.70  \\
\hline
\end{tabular}
\vspace{-10pt}
\end{table}

\subsection{Baselines}
The baselines used in our experiments are summarized in Table~\ref{tab:baselines}. While some baselines appear conceptually simple, recent work has shown that they are often able to outperform very recent neural approaches for session-based recommendation tasks
\citep{jannach2017recurrent,ludewig2018evaluation,ludewig2019performance}. Unlike neural methods like \emph{GRU4REC}, these methods can be continuously updated over time to take newly published articles into account. A comparison of \emph{GRU4REC} with some of our baselines in a streaming scenario is provided in \cite{jugovac2018streamingrec}, and specifically in the news domain in \cite{moreira2018news}, which is why we do not include \emph{GRU4REC} and similar methods here.

\begin{table}[!htbp]
\centering
\vspace{-3pt}
\caption{Baseline recommendation algorithms.}
\label{tab:baselines}
\vspace{-5pt}
\footnotesize
\begin{tabular}{p{2cm}p{5.3cm}}
\hline
\multicolumn{2}{p{6cm}}{Association Rules-based and Neighborhood Methods}\\ \hline
\emph{Co-Occurrence (CO)} &  Recommends articles commonly viewed together with the last read article in previous user sessions \cite{ludewig2018evaluation,jugovac2018streamingrec}. \\
\emph{Sequential Rules (SR)} &  The method also uses association rules of size two. It however considers the sequence of the items within a session and uses a weighting function when two items do not immediately appear after each other \citep{ludewig2018evaluation}.  \\
\emph{Item-kNN} & Returns the most similar items to the last read article using the cosine similarity between their vectors of co-occurrence with other items within sessions. This method has been commonly used as a baseline for neural approaches, e.g., in  \citep{hidasi2016}.\tablefootnote{We also made experiments with session-based methods proposed in \cite{ludewig2018evaluation} (e.g. V-SkNN), but they did not lead to results that were better than the \emph{SR} and \emph{CO} methods.}\\
\hline
\multicolumn{2}{l}{Non-personalized Methods}\\ \hline
\emph{Recently Popular (RP)} & This method recommends the most viewed articles within a defined set of recently observed user interactions on the news portal (e.g., clicks during the last  hour). Such a strategy proved to be very effective in the \emph{2017 CLEF NewsREEL Challenge} \cite{ludmann2017recommending}. \\
\emph{Content-Based (CB)} &  For each article read by the user, this method suggests recommendable articles with similar content to the last clicked article, based on the cosine similarity of their \emph{Article Content Embeddings} (generated by the \emph{CNN} technique described in Table~\ref{tab:feature-extraction}). \\ \hline
\end{tabular}
\vspace{0pt}
\end{table}

\paragraph{Replicability}
We publish the data and source code used in our experiments 
online\footnote{\url{https://github.com/gabrielspmoreira/chameleon\_recsys}}, including the code for \emph{CHAMELEON}, which is implemented using \emph{TensorFlow}. 

\section{Experimental Results}
The results for the \emph{G1} and \emph{Adressa} datasets after (hyper-)parameter optimization for all methods are presented\footnote{The highest values for a given metric are highlighted in bold. The best values for the \emph{CHAMELEON} configurations are printed in italics. If the best results are significantly
different ($ p < 0.001 $) from all other algorithms, they are marked with *. We used paired Student's t-tests with Bonferroni correction for significance
tests.} in Tables~\ref{tab:g1_content_results} and~\ref{tab:adressa_content_results}.



\vspace{-5pt}
\begin{table}[h!t]
\centering
\caption{Results for the G1 dataset.}
\label{tab:g1_content_results}
\vspace{-5pt}
\footnotesize
\begin{tabular}{lllll}
\hline
 \emph{Recommender}  & \emph{HR@10}  & \emph{MRR@10}  & \emph{COV@10}  & \emph{ESI-R@10} \\
\hline
\multicolumn{5}{p{6cm}}{\emph{CHAMELEON with ACEs generated differently}} \\ \hline \hline

\emph{No-ACE} & 0.6281 & 0.3066 & 0.6429 & 6.3169  \\ \hline
\emph{CNN} & 0.6585 & 0.3395 & \textit{0.6493} & 6.2874 \\
\emph{GRU} & 0.6585 & 0.3388 & 0.6484 & 6.2674 \\ \hline
\emph{W2V*TF-IDF} & 0.6575 & 0.3291 & 0.6500 & 6.4187 \\
\emph{LSA} & \textbf{0.6686}* & \textbf{0.3423} & 0.6452 & 6.3833 \\
doc2vec & 0.6368 & 0.3119 & 0.6431 & \textit{6.4345} \\
\hline
\multicolumn{5}{p{6cm}}{\emph{Baselines}} \\ \hline \hline
\emph{SR} & 0.5911 & 0.2889 & 0.2757 & 5.9743 \\
\emph{Item-kNN} & 0.5707 & 0.2801 & 0.3892 & 6.5898 \\
\emph{CO} & 0.5699 & 0.2625 & 0.2496 & 5.5716 \\
\emph{RP} & 0.4580 & 0.1994 & 0.0220 & 4.4904 \\
\emph{CB} & 0.3703 & 0.1746 & \textbf{0.6855}* & \textbf{8.1683}* \\
\hline  \hline
\end{tabular}
\vspace{0pt}
\end{table}

\vspace{-5pt}
\begin{table}[h!t]
\centering
\caption{Results for the Adressa dataset.}
\vspace{-5pt}
\footnotesize
\label{tab:adressa_content_results}
\begin{tabular}{lllll}
\hline
 \emph{Recommender}  & \emph{HR@10}  & \emph{MRR@10}  & \emph{COV@10}  & \emph{ESI-R@10} \\
\hline
\multicolumn{5}{p{6cm}}{\emph{CHAMELEON with ACEs generated differently}} \\ \hline \hline

\emph{No-ACE} & 0.6816 & 0.3252 & \textit{0.8185} & 5.2453  \\ \hline
\emph{CNN} & 0.6860 & 0.3333 & 0.8103 & 5.2924  \\
\emph{GRU} & 0.6856 & 0.3327 & 0.8096 & 5.2861  \\ \hline
\emph{W2V*TF-IDF} & 0.6913 & 0.3402 & 0.7976 & 5.3273  \\
\emph{LSA} & \textbf{0.6935} & \textbf{0.3403} & 0.8013 & 5.3347  \\
\emph{doc2vec} & 0.6898 & 0.3402 & 0.7968 & \textit{5.3417}  \\
\hline
\multicolumn{5}{p{6cm}}{\emph{Baselines}} \\ \hline \hline
\emph{SR} & 0.6285 & 0.3020 & 0.4597 & 5.4445 \\
\emph{Item-kNN} & 0.6136 & 0.2769 & 0.5287 & 5.4668 \\
\emph{CO} & 0.6178 & 0.2819 & 0.4198 & 5.0785 \\
\emph{RP} & 0.5647 & 0.2481 & 0.0542 & 4.1464 \\
\emph{CB} & 0.3273 & 0.1197 & \textbf{0.8807}* & \textbf{7.6534}* \\
\hline  \hline
\end{tabular}
\vspace{-10pt}
\end{table}

\paragraph{Accuracy Results.} In general, we can observe that considering content information is in fact highly beneficial in terms of recommendation accuracy. It is also possible to see that the choice of the article representation matters. Surprisingly, the long-established \emph{LSA} method was the best performing technique to represent the content for both datasets in terms of accuracy, even when compared to more recent techniques using pre-trained word embeddings, such as the \emph{CNN} and \emph{GRU}.

For the \emph{G1} dataset, the \emph{Hit Rates} (\emph{HR}) were improved by around 7\% and the \emph{MRR} by almost 12\% when using the LSA representation instead of the \emph{No-ACE setting}. For the \emph{Adressa} dataset, the difference between the \emph{No-ACE} settings and the hybrid methods leveraging text are less pronounced. The improvement using \emph{LSA} compared to the \emph{No-ACE setting} was around 2\% for \emph{HR} and 5\% for \emph{MRR}.

Furthermore, for the \emph{Adressa} dataset, it is possible to observe that all the \emph{unsupervised} methods (\emph{LSA}, \emph{W2V*TF-IDF}, and \emph{doc2vec}) for generating \emph{ACEs} performed better than the \emph{supervised} ones, differently from the \emph{G1} dataset.
A possible explanation can be that the \emph{supervised} methods depend more on the \emph{quality} and \emph{depth} of the available article metadata information. 
While the \emph{G1} dataset uses a fine-grained categorization scheme (461 categories), the categorization of the \emph{Adressa} dataset is much more coarse (41 categories).

Among the baselines, \emph{SR} leads to the best accuracy results, but does not match the performance of the content-agnostic \emph{No-ACE} settings for an RNN. This indicates that the hybrid approach of considering additional contextual information, as done by \emph{CHAMELEON}'s \emph{NAR} module in this condition, is important.

Recommending only based on content information (\emph{CB}), as expected, does not lead to competitive accuracy results, because the popularity of the items is not taken into account (which \emph{SR} and neighborhood-based methods implicitly do). Recommending only recently popular articles (\emph{RP}) works better than \emph{CB}, but does not match the performance of the other methods.

\paragraph{Coverage and Novelty.} In terms of coverage (\emph{COV@10}), the simple \emph{Content-Based (CB)} method leads to the highest value, as it recommends across the entire spectrum based solely on content similarity, without considering the popularity of the items. It is followed by the various \emph{CHAMELEON} instantiations, where it turned out that the specifically chosen content representation is not too important in this respect.


As expected, the \emph{CB} method also frequently recommends long-tail items, which also leads to the highest value in terms of novelty (\emph{ESI-R@10}). The popularity-based method (\emph{RP}), in contrast, leads to the lowest novelty value. From the other methods, the traditional \emph{Item-KNN} method, to some surprise, leads to the best novelty results, even though neighborhood-based methods have a certain popularity bias. Looking at the other configurations, using \emph{unsupervised} methods to represent the text of the articles can help to drive the recommendations a bit away from the popular ones.

\section{Summary and Conclusion}
The consideration of content information for news recommendation proved to be important in the past, and therefore many hybrid systems were proposed in the literature. In this work, we investigated the relative importance of incorporating content information in both streaming- and session-based recommendation scenarios. Our experiments highlighted the value of content information by showing that it helped to outperform otherwise competitive baselines. Furthermore, the experiments also demonstrated that the choice of the article representation can matter. However, the value of considering additional content information in the process depends on the quality and depth of the available data, especially for \emph{supervised} methods. From a practical perspective, this indicates that quality assurance and curation of the content information can be essential to obtain better results.

\balance
\bibliographystyle{ACM-Reference-Format}
\bibliography{ms}

\end{document}